\documentclass[twocolumn]{article}
\usepackage{times}
\usepackage{soul}
\usepackage{url}
\usepackage[hidelinks]{hyperref}
\usepackage[utf8]{inputenc}
\usepackage[small]{caption}
\usepackage{graphicx}
\usepackage{amsmath}
\usepackage{amsthm}
\usepackage{booktabs}
\usepackage{algorithm}
\usepackage{algorithmic}
\usepackage[switch]{lineno}
\usepackage[acronym, shortcuts]{glossaries}
\usepackage{subcaption}
\usepackage{multicol}
\usepackage{multirow}
\usepackage{color}
\usepackage{paralist}

\glsdisablehyper

\newacronym{DNN}{DNN}{deep neural network}
\newacronym{MLP}{MLP}{multilayer perceptron}
\newacronym{ReLU}{ReLU}{rectified linear unit}
\newacronym{AE}{AE}{autoencoder}
\newacronym{DL}{DL}{deep learning}
\newacronym{AI}{AI}{artificial intelligence}
\newacronym{ML}{ML}{machine learning}
\newacronym{NN}{NN}{neural network}
\newacronym{FL}{FL}{federated learning}
\newacronym{SGD}{SGD}{stochastic gradient descent}

\newacronym{STDP}{STDP}{spike timing-dependent plasticity}
\newacronym{SNN}{SNN}{spiking neural network}
\newacronym{DVS}{DVS}{dynamic vision sensor}
\newacronym{LIF}{LIF}{leaky integrate-and-fire}
\newacronym{ATIS}{ATIS}{asynchronous time-based image sensor}

\newacronym{LR}{LR}{learning rate}
\newacronym{MSE}{MSE}{mean squared error}
\newacronym{ASR}{ASR}{attack success rate}
\newacronym{SSIM}{SSIM}{structural similarity index}
\newacronym{SVD}{SVD}{singular value decomposition}

\newacronym{ABS}{ABS}{artificial brain stimulation}
\newacronym{STRIP}{STRIP}{strong intentional perturbation}
\newacronym{NC}{NC}{neural cleanses}

\DeclareMathOperator*{\argmin}{arg\,min}

\title{Time-Distributed Backdoor Attacks on Federated Spiking Learning}

\author{
{\rm Gorka Abad}\\
Radboud University\\
Ikerlan Research Centre
\and
{\rm Stjepan Picek}\\
Radboud University\\
Delft University of Technology
\and
{\rm Aitor Urbieta}\\
Ikerlan Research Centre
} 

\begin{document}

\maketitle

\begin{abstract}    
    This paper investigates the vulnerability of spiking neural networks (SNNs) and federated learning (FL) to backdoor attacks using neuromorphic data. Despite the efficiency of SNNs and the privacy advantages of FL, particularly in low-powered devices, we demonstrate that these systems are susceptible to such attacks. We first assess the viability of using FL with SNNs using neuromorphic data, showing its potential usage. Then, we evaluate the transferability of known FL attack methods to SNNs, finding that these lead to suboptimal attack performance. Therefore, we explore backdoor attacks involving single and multiple attackers to improve the attack performance. Our primary contribution is developing a novel attack strategy tailored to SNNs and FL, which distributes the backdoor trigger temporally and across malicious devices, enhancing the attack's effectiveness and stealthiness. In the best case, we achieve a 100\% attack success rate, 0.13 MSE, and 98.9 SSIM. Moreover, we adapt and evaluate an existing defense against backdoor attacks, revealing its inadequacy in protecting SNNs. This study underscores the need for robust security measures in deploying SNNs and FL, particularly in the context of backdoor attacks.
\end{abstract}

\section{Introduction}
\label{sec:introduction}

\Acp{DNN} achieve outstanding performance on numerous tasks~\cite{Goodfellow-et-al-2016}, such as computer vision, speech recognition, or natural language processing. \Acp{DNN} leverage their full potential by learning from a large corpus of data. The data is processed in an iterative learning process, which requires powerful computing devices, i.e., GPUs, takes much time, and involves significant training costs~\cite{dhar2020carbon}.

In this context, \acp{SNN} are highly useful. \acp{SNN} are an emerging field that leverages neuromorphic computing to enable energy-efficient computing. \acp{SNN} use neuromorphic data, time encoding the information, thus being much more compressed than regular data, e.g., images. \acp{SNN} perform similarly to the \emph{traditional} \ac{ML} models while drastically reducing their energy consumption.

At the same time, privacy regulations have led the research community to investigate privacy-preserving \ac{ML} schemes, such as \ac{FL}. \ac{FL} is a training scheme that enables data decentralization, where numerous devices or clients aim to train a \ac{ML} model without sharing private data. \ac{FL} has been extensively evaluated using non-neuromorphic data, including \acp{SNN}~\cite{wang2023efficient,venkatesha2021federated}, showcasing its usability. Research has yet to explore the viability of combining \ac{FL} with neuromorphic data.

Due to the increasing attention \ac{ML} has received, the research community has been focusing on evaluating its security. In the last few years, new threats have been discovered~\cite{he2020towards}, such as adversarial examples, model inversion, and backdoor attacks, to name a few. Regarding \acp{SNN}, recent investigations have concluded that \acp{SNN} are also vulnerable to some of these attacks, i.e., adversarial examples~\cite{sharmin2019comprehensive,xu2022securing,buchel2021adversarial} and backdoor attacks~\cite{abad2023sneaky,abad2022poster}.
In the context of \ac{FL}, security and privacy evaluations have also been in the scope of security experts~\cite{abad2021security,mothukuri2021survey}, concluding that \ac{FL} is vulnerable to privacy attacks, such as membership inference, and security attacks, such as backdoor attacks. 

This investigation tackles some important research gaps: the lack of i) evaluation of \ac{FL} with \acp{SNN} and neuromorphic data and ii) the security analysis of \ac{FL} with \acp{SNN} and neuromorphic data. In this paper, we are the first to investigate how to train an \ac{FL} setup with \acp{SNN} and then evaluate if the existing \ac{FL} attacks also apply (after adaptation) to neuromorphic data and \acp{SNN}. Furthermore, we develop a novel attack vector unique to \ac{FL} with \acp{SNN} and neuromorphic data, which splits the backdoor trigger in time, named \emph{Time Bandits}, showcasing some new challenges and the necessity of stronger defenses.

In summary, our contributions are summarized as:
\begin{compactenum}
    \item We are the first to evaluate the viability of \acp{SNN} in \ac{FL} and neuromorphic data in various settings.
    \item We pioneer the investigation into backdoor attacks in \ac{FL} using \acp{SNN} and neuromorphic data. More precisely, we:
    \begin{compactitem}
        \item Adapt known backdoor attacks from \ac{FL} in non-neuromorphic scenarios.
        \item Design a novel backdoor trigger specific to \ac{FL} scenarios using neuromorphic data, obtaining the novel Time Bandits attack.
    \end{compactitem}
    \item We consider single and multiple attacker scenarios and further evaluate the stealthiness of the proposed triggers.
    \item We investigate if an already known backdoor defense in standard \ac{FL} is applicable under neuromorphic data. 
\end{compactenum}


\section{Background}
\label{sec:background}

\subsection{SNNs \& Neuromorphic Data}

\begin{figure}[htb]
    \centering
    \begin{subfigure}[b]{0.24\linewidth}
        \centering
        \includegraphics[width=\textwidth]{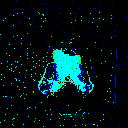}
    \end{subfigure}
    \hfill  
    \begin{subfigure}[b]{0.24\linewidth}
        \centering
        \includegraphics[width=\textwidth]{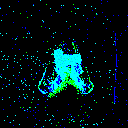}
    \end{subfigure}
    \hfill
    \begin{subfigure}[b]{0.24\linewidth}
        \centering
        \includegraphics[width=\textwidth]{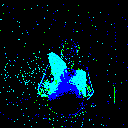}
    \end{subfigure}
    \hfill
    \begin{subfigure}[b]{0.24\linewidth}
        \centering
        \includegraphics[width=\textwidth]{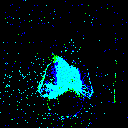}
    \end{subfigure}
    \caption{Neuromorphic data at different frames.}
    \label{fig:gesture}
\end{figure}

Neuromorphic data is an emerging approach to encoding information inspired by biological neural systems. Contrary to regular data representation, e.g., images, neuromorphic data exploit the temporal dynamics and encode the information in an event-driven and asynchronous way. This encoding finds its application in numerous fields, with a notable emphasis on visual processing; see~\autoref{fig:gesture}.

\Acp{DVS} cameras and \acp{SNN} enable capturing and processing neuromorphic data. \Acp{DVS} cameras capture the changes in luminance through time as a stream of events, capturing real-time and efficient visual scene representation. \Acp{SNN} leverage neuromorphic data, enabling efficient data processing through discrete events or spikes.

Formally, let $I(\omega)$ represent the neuromorphic data input at time $\omega$, where $I(\omega)$ is a stream of events capturing changes in luminance. The \ac{SNN} processes this input through its layers of spiking neurons. The behavior of a spiking neuron $j$ at time $\omega$ is given by the \ac{LIF} model:

\begin{align}
\label{eq:LIF}
    h_j(\omega) = 
    \begin{cases}
        1, & \text{if } V_j(\omega) \geq \nu_j, \\
        0, & \text{otherwise},
    \end{cases}
\end{align}
where $V_j(\omega)$ is the membrane potential of neuron $j$ at time $\omega$, and $\nu_j$ is the firing threshold. The membrane potential evolves based on the input spikes and is subject to reset after a spike is emitted. 
Note that~\autoref{eq:LIF} is non-differentiable, disabling the usage of calculating gradients and, thus, performing backpropagation. This non-differentiable behavior necessitates specialized training approaches such as \ac{STDP} or surrogate gradients. Surrogate training allows the approximation of the derivatives to perform backpropagation, enabling compatibility with Adam~\cite{kingma2014adam} or \ac{SGD}~\cite{amari1993backpropagation}. The latter achieves better performance~\cite{lee2016training}, representing the method we follow in our work.

\subsection{Federated Learning}

\Ac{FL}~\cite{mcmahan2017communication} is a decentralized \ac{ML} technique that enables model training across multiple devices while keeping the data localized. In centralized \ac{ML} approaches, data is typically centralized, posing privacy concerns. \Ac{FL} addresses these concerns by allowing the training of a global model collaboratively across distributed devices without exchanging raw data.

Let $\mathcal{D}_1, \mathcal{D}_2, \ldots, \mathcal{D}_n$ represent the local datasets available on $n$ different devices. Each local dataset $\mathcal{D}_i$ is associated with a model parameterized by $\theta_i$. \Ac{FL} aims to learn a global model parameterized by $\Theta$ that captures the knowledge from all participating devices. This global model is updated collaboratively through an iterative process.
Each iteration selects a subset of devices to participate in the model update. Let $\mathcal{B}_t \subset \{1, 2, \ldots, n\}$ represent the set of devices selected at iteration $t$. The global model, $\Theta$, is updated by aggregating the local updates from the selected devices:

\begin{align*}
    \Theta_{t+1} = \frac{1}{|\mathcal{B}_t|} \sum_{i \in \mathcal{B}_t} \theta_{i, t+1},
\end{align*}
where $\theta_{i, t+1}$ is the local model update from device $i$ at iteration $t+1$. The aggregation process ensures that the global model benefits from the knowledge of all participating devices while preserving the privacy of individual data.

\subsection{Backdoor Attacks}

Backdoor attacks alter a model's behavior during training, causing it to exhibit abnormal behavior at test time~\cite{gu2019badnets}. A backdoored model incorrectly classifies inputs with a trigger while behaving correctly on clean inputs. The training set is adjusted to include malicious data samples containing the trigger in the context of data poisoning backdoors. In the image domain, the trigger may manifest as a pixel pattern in a specific part of an image. When the algorithm is trained on a combination of clean and backdoor data, the model learns explicitly to misclassify inputs containing the pixel pattern (i.e., the trigger) to a predefined target label.

Formally, an algorithm $f_\theta(\cdot)$ is trained on a mixed dataset comprising both clean and backdoor data, with the mixing ratio controlled by $\epsilon = \frac{m}{n}$, where $n$ is the size of the clean dataset, $m$ is the size of the backdoor dataset, and $m \ll n$. The backdoor dataset $\mathcal{D}_{bk}$ consists of backdoor samples $\{(\hat{\mathbf{x}}, \hat{y})\}^m \in \mathcal{D}_{bk}$, where $\hat{\mathbf{x}}$ is the sample containing the trigger, and $\hat{y}$ is the target label. For a clean dataset of size $n$, the training procedure seeks to find $\theta$ by minimizing the loss function $\mathcal{L}$:
\begin{align}
\footnotesize
    \label{eq:train}
    \theta' = \argmin_\theta \sum_{i=0}^n \mathcal{L}(f_\theta(\mathbf{x}_i),y_i),
\end{align}
where $\mathbf{x}$ is the input, and $y$ is the ground truth label. During training with backdoor data,~\autoref{eq:train} is modified to incorporate the backdoor behavior, expressed as:
\begin{align*}
\footnotesize
    \theta' = \argmin_\theta \sum_{i=0}^n \mathcal{L}(f_\theta(\mathbf{x}_i),y_i) + \sum_{j=0}^m \mathcal{L}(f_\theta(\hat{\mathbf{x}}_j),\hat{y}_j).
\end{align*}

In \ac{FL}, backdoor attacks are also prominent~\cite{Bagdasaryan2020}. A malicious device includes poisoned data samples in the local dataset and injects them into the local model through training, following the same approach explained above. When the number of devices in the network is large, the backdoor effect may vanish during aggregation---a single malicious model out of many will have a low impact due to the averaging process. The local model is scaled by scalar $\gamma = \frac{n}{\hat{n}}$, where $\hat{n}$ is the number of malicious devices, impacting the global model more~\cite{Bagdasaryan2020}.

\section{Threat Model}
\label{sec:threat}

We consider the same threat model as in prior studies~\cite{Bagdasaryan2020}. Different devices aim to train a shared model for solving a specific task. We consider all the devices benign except one (or a small subset of non-colluding malicious devices), which aims to inject a backdoor into the global model. We assume that the attacker can fully access the local model and dataset. The attacker can also control its local training procedure and select arbitrary hyperparameters as the learning rate or the number of epochs. The attacker can also modify the weights of the local model before sharing it with the server. However, no one can directly interact with other devices or access others' private data. The attacker aims to inject a backdoor using a data poisoning-based, dirty label methodology~\cite{gu2019badnets}.
We also limit our research to the digital image domain, where triggers are intended to be injected in digital samples rather than applied physically in the wild~\cite{bagdasaryan2021blind}. Regarding the \ac{FL} type, we consider horizontal federated learning, where different devices contain distinct samples for the same set of features, which is usually considered in recent work~\cite{Bagdasaryan2020}.

\section{Methodology}
\label{sec:methodology}

This section gives a detailed explanation and rationale for the considered attacks. We propose an attack considering a single malicious device and multiple non-colluding malicious devices, i.e., \emph{Time Bandits}.

\subsection{Single Attacker}

\begin{figure}[htb]
    \centering
     \includegraphics[width=\linewidth]{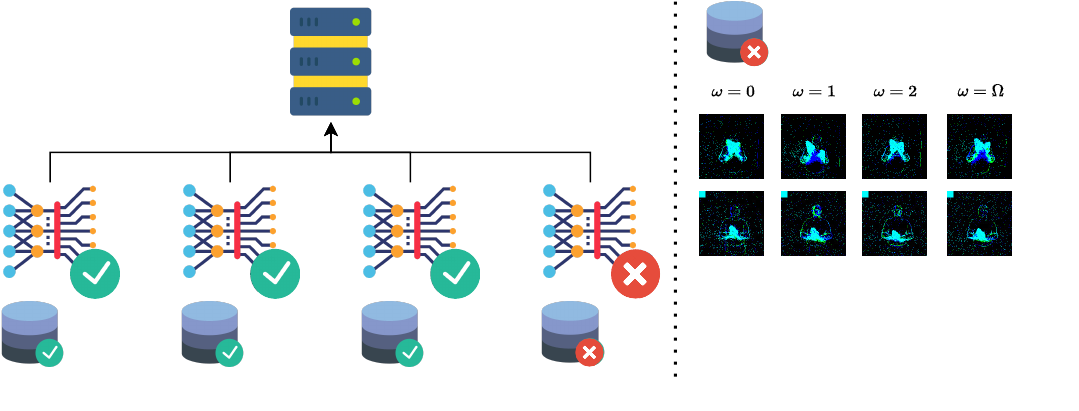}
    \caption{Overview of the single attacker backdoor attack.}
    \label{fig:single-attacker}
\end{figure}

In a single attacker scenario, the attacker aims to introduce a backdoor into the global model by leveraging its private data and model (see~\autoref{fig:single-attacker}). We adopt the \emph{vanilla} backdoor attack in \ac{FL}~\cite{Bagdasaryan2020} to achieve this, and we adapt it to work with neuromorphic data. Specifically, the attacker chooses a location, trigger size, and trigger polarity. Instead of evaluating these parameters exhaustively, we rely on prior research to select an already used (and well-performing) configuration~\cite{abad2023sneaky}. Specifically, we employ a trigger size equivalent to 30\% of the image, placing the trigger in the top-left corner with a polarity of 1 (light blue).

The attacker incorporates images containing the trigger into the local dataset. The number of poisoned samples is controlled by $\epsilon$. The trigger is consistently placed in the exact location across all frames $\Omega$, and the polarity remains constant. Following local training, the attacker scales the malicious model $\hat{\theta}$ and transmits it to the server for aggregation.
Note that the attacker may not be selected in every iteration; its probability will depend on the number of devices selected, which may affect the attack performance. In later sections, we test this hyperparameter; see~\autoref{subsec:multiple-attackers}.

\subsection{Multiple Attackers - Time Bandits}

\begin{figure}[htb]
    \centering
    \includegraphics[width=\linewidth]{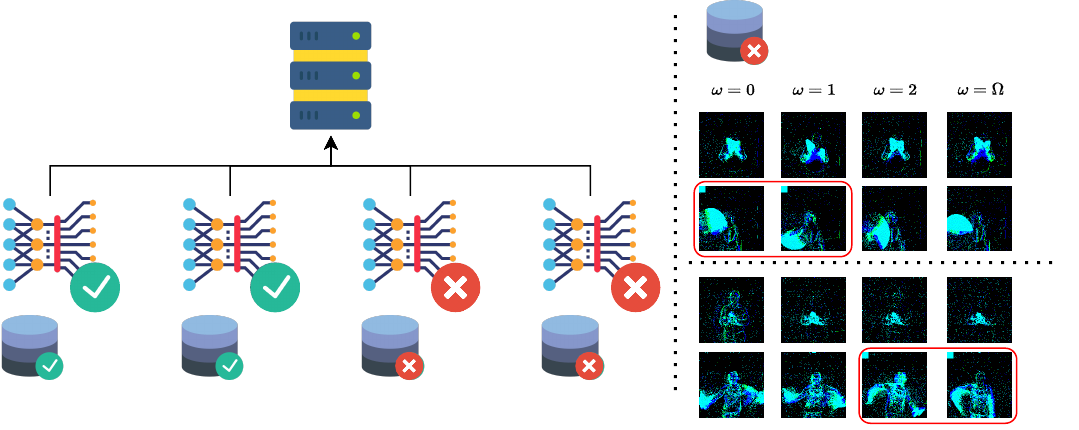}
    \caption{Overview of Time Bandits with two malicious devices. One of the attackers contaminates the first $\Omega/2$ frames, while the other poisons the last $\Omega/2$ frames.}
    \label{fig:multiple-attackers}
\end{figure}

In a multiple attacker scenario, multiple devices collaborate toward a single goal without colluding (see~\autoref{fig:multiple-attackers}). Inspired by the work of Xie et al.~\cite{xie2020dba}, where a distributed backdoor attack in \ac{FL} involves multiple malicious devices splitting the trigger into smaller parts, we propose a novel attack leveraging unique features of neuromorphic data such as its dynamism, where the trigger is split in time. We denote this attack as \emph{Time Bandits}.

Instead of splitting the trigger spatially, we propose splitting it through time. Each device injects the trigger into a specific subset of frames. For instance, with two malicious devices and data samples spanning $\Omega = 16$ frames, the first malicious device injects the trigger in the first eight frames. In contrast, the second malicious device affects the last eight frames. The choice of this strategy may vary and is not fixed to this example. We exemplify with two attackers; however, the attacker can be generalized to an arbitrary number of attackers. More than two attackers are considered in~\autoref{subsec:multiple-attackers}. At test time, a data sample with the trigger in \textbf{all} frames launches the backdoor effect.

Formally, let $N$ denote the total number of devices in the \ac{FL} system, and let $\mathcal{D}_i$ represent the local device dataset $i$. For a given data sample $\mathbf{x}$ consisting of a sequence of frames $\mathbf{x} = {\mathbf{x}_1, \mathbf{x}_2, \ldots, \mathbf{x}_\Omega}$, where $\Omega$ is the total number of frames in the sample, we introduce a binary function $h_i(\omega)$ for device $i$, representing whether the trigger is injected into the frame $\omega$. The function $h_i(\omega)$ takes the value of 1 if the trigger is injected and 0 otherwise.
The local model update for each malicious device $i$ at iteration $t+1$ is then given by minimizing the loss function~$\mathcal{L}$:
\begin{align*}
    \begin{split}
    \theta_{\text{attacker}i, t+1} = & \argmin_\theta \sum_{i=0}^n \mathcal{L}(f_\theta(\mathbf{x}_i),y_i) +\\ 
    & \quad \sum_{j=0}^m \mathcal{L}(f_\theta(\hat{\mathbf{x}}_j),\hat{y}_j),    
    \end{split}
\end{align*}
where $\mathbf{x}_i$ represents a clean input, $y_i$ is the corresponding label, $\hat{\mathbf{x}}_j$ is a backdoor sample, and $\hat{y}_j$ is the target label.

The distributed backdoor injection is achieved through the binary function $h_i(\omega)$. Specifically, for each frame $\omega$:
\begin{align*}
B_i(\omega) =
\begin{cases}
1, & \text{\small if $\omega \leq \frac{\Omega}{N}$, device $i$ is the first attacker}, \\
1, & \text{\small if $\frac{\Omega}{N} < \omega \leq \frac{2\Omega}{N}$, device $i$ is the 2\textsuperscript{nd} attacker}, \\
\vdots \\
1, & \text{\small if $\frac{(N-1)\Omega}{N} < \omega \leq \Omega$, $i$ is the last attacker}, \\
0, & \text{\small otherwise}.
\end{cases}
\end{align*}

At test time, a data sample with the trigger injected in all frames is represented by $\sum_{i=1}^{N} h_i(\omega)$, and this launches the backdoor effect in the \ac{FL} system.

Using neuromorphic data opens new challenges and possibilities for backdoor triggers. This work is the first to explore splitting triggers in time and distributing them among attackers. This novel approach opens a new avenue for other attacks, such as triggers that change shape, position, or polarity through time, which we aim to explore in future work.

\section{Experimentation}

\subsection{Experimental Setup}


\textbf{Datasets.} We use three datasets: N-MNIST~\cite{orchard2015converting}, CIFAR10-DVS~\cite{li2017cifar10} (N-CIFAR10), and DVS128-Gesture~\cite{amir2017low} (Gesture). We use N-MNIST and N-CIFAR10 since their non-neuromorphic versions are common benchmarking datasets in computer vision for security/privacy in \ac{ML}.
N-MNIST is a spiking version of MNIST~\cite{lecun1998mnist}, which contains $34\times34$ 60\,000 training, and 10\,000 test samples. An \ac{ATIS}~\cite{posch2010high} captured the dataset across the 10 MNIST digits on an LCD monitor. The N-CIFAR10 dataset is the spiking version of the CIFAR10~\cite{krizhevsky2009learning} dataset, which contains 9\,000 training, and 1\,000 test $128\times128$ samples, corresponding to 10 classes. Lastly, the Gesture dataset is a ``fully neuromorphic'' dataset created for \acp{SNN} tasks. The Gesture dataset collects real-time motion captures from 29 subjects making 11 different hand gestures under three illumination conditions, creating 1\,176 $128\times128$ training samples and 288 test samples.
For all datasets, the samples' shape is $\Omega\times P\times H\times W$, where $\Omega$ is the number of time steps (we set it to $\Omega=16$ based on~\cite{SpikingJelly}), $P$ is the polarity, $H$ is the height, and $W$ is the width.

\textbf{Models.} We consider three network architectures for the victim classifiers used in related works~\cite{fang2021incorporating}. The N-MNIST dataset's network comprises a single convolutional layer and a fully connected layer. For the N-CIFAR10 dataset, the network contains two convolutional layers followed by batch normalization and max pooling layers. Then, two fully connected layers with dropout are added, and lastly, a voting layer---for improving the classification robustness~\cite{fang2021incorporating}---of size ten is incorporated. Finally, five convolutional layers with batch normalization and max pooling for the Gesture dataset, two fully connected layers with dropout, and a voting layer compose the network.

We first evaluate model performance on clean data, serving as a baseline for later experiments. We split the data independently and identically distributed (IID) for our experiments. We consider non-IID data in~\autoref{sec:non-iid}. \autoref{tab:baseline} shows the clean accuracy for different datasets with clean data using a single device. All the results are the averaged values over four execution rounds.

\begin{table}[htb]
\centering
\caption{Baseline clean accuracy (\%) with a single device.}
\label{tab:baseline}
\resizebox{0.7\columnwidth}{!}{%
\begin{tabular}{ccc}
\toprule
Dataset   & Epochs & Accuracy (\%) \\ \midrule
N-MNIST   &   10   &     99.27     \\
N-CIFAR10 &   30   &     67.60      \\
Gesture   &   64   &     91.67     \\ \bottomrule
\end{tabular}%
}
\end{table}

Interestingly, when using the federated setup with multiple devices, we observe a notable accuracy degradation, see~\autoref{tab:baseline_fl}. The accuracy degradation is mainly observed on complex datasets such as Gesture and N-CIFAR10. On Gesture, the degradation is more significant due to the small size of the dataset; each device contains few samples, which makes the model harder to generalize. Likewise, the degradation is most significant with many devices in the network and when the number of selected devices is small.

\begin{table}[htb]
\centering
\caption{Clean accuracy (\%) baseline for different FL setups.}
\label{tab:baseline_fl}
\resizebox{\columnwidth}{!}{%
\begin{tabular}{@{}cccccccccc@{}}
\toprule
       & \multicolumn{3}{c}{10} & \multicolumn{3}{c}{25} & \multicolumn{3}{c}{50} \\
 & 0.1    & 0.5    & 1    & 0.1    & 0.5    & 1    & 0.1    & 0.5    & 1    \\ \midrule
N-MNIST     &     97.93   &  98.18      &   98.2   &   96.29     &   96.6     &   96.72   &   93.25     &  93.71       &  94.01    \\
N-CIFAR10   &   49.7     &    54.5    &   53.3   &   43.7     &    43.3    &   41.3   &   27.4     &    30.4    & 53.3    \\
Gesture     &    70.48    &    75.69    &  71.87   &   50.69     &    62.5    &   59.37   &   47.92     &  44.45      &   50.35   \\ \bottomrule
\end{tabular}%
}
\end{table}

\subsection{Single Attacker}

In the single attacker scenario, the attacker crafts poisoned data and includes them in the local dataset, so the backdoor is injected into the model through local training. 
We observe that without scaling, the attack results are ineffective; see~\autoref {tab:single_attk}. The attack achieves a 39.59\% \ac{ASR} with ten devices and a 9.74\% with 50 devices for N-MNIST. As expected, the more benign the devices, the worse the attack performance. The same effect is also noticeable for the other datasets. Therefore, we also evaluate the backdoor performance under model scaling. The attacker scales the model by a scalar based on the number of total devices on the network. The attack performance increases in all the settings (see~\autoref{tab:single_attk}); however, in most cases, the performance is not as good as the state-of-the-art~\cite{Bagdasaryan2020}.

\begin{table*}[!htb]
\centering
\caption{ASR (\%) of a single attacker backdoor attack under different settings. Without and with scaling. $\epsilon = 0.1$.}
\label{tab:single_attk}
\resizebox{\textwidth}{!}{%
\begin{tabular}{cccccccccc}
\toprule
          & \multicolumn{3}{c}{10} & \multicolumn{3}{c}{25} & \multicolumn{3}{c}{50} \\
          & 0.1    & 0.5    & 1    & 0.1    & 0.5    & 1    & 0.1    & 0.5    & 1    \\ \midrule
N-MNIST   & 38.59 / 39.98 & 38.39 / 100 & 30.38 / 100 & 9.73 / 39.41 & 10.09 / 100 & 10.66 / 100 & 9.31 / 39.36 & 9.48 / 100 & 9.74 / 100 \\
N-CIFAR10 & 60.27 / 63.43 & 44.97 / 97.63 & 29.60 / 99.60 & 12.57 / 14.00 & 11.20 / 94.93 & 10.83 / 97.83 & 15.90 / 31.40 & 10.67 / 25.70 & 9.27 / 48.53 \\ 
Gesture   &  19.91 / 5.90 & 7.06 / 7.75 & 7.75 / 17.59 & 3.70 / 11.23 & 6.60 / 29.51 & 7.52 / 24.77 & 5.67 / 24.65 & 6.37 / 4.75 & 6.94 / 16.44 \\ \bottomrule
\end{tabular}%
}
\end{table*}

Attacking \ac{FL} with \acp{SNN} is generally more complex than attacking its non-neuromorphic counterpart due to the lower performance of \acp{SNN}. However, the attack achieves top backdoor performance on less complex datasets such as N-MNIST, where the samples are simple and without noise, i.e., 100\% \ac{ASR}. It is important to remark on the importance of the rounds at which the attacker is selected. For instance, if the attacker is only selected in an early round, the backdoor effect will not be retained and will vanish. This will also be important in later experiments.

\subsection{Multiple Attackers}

Under a multiple attacker setup, we use our novel attack Time Bandits; we consider various scenarios where different devices are compromised and non-colluding; they collaborate to achieve the same goal. Precisely, we begin considering two malicious devices, where, in a coordinated manner, they split the trigger through time and include it in their respective dataset (more malicious devices are considered in~\autoref{subsec:multiple-attackers}).

Note that two malicious devices comprise 20\%, 8\%, and 4\% of the total federated network for 10, 25, and 50 devices. Compared with the single attacker scenario, we observe a significant increase in the performance of the attack (see~\autoref{tab:multiple_attk_2}). Precisely, for N-MNIST and N-CIFAR10, the attack achieves top performance. However, for the Gesture dataset, we observe the same problem as stated before due to the small size of the dataset. Recall that in a multiple attacker setup, the trigger is split into multiple parties, and the malicious samples are different per device, i.e., attacker 1 contains different samples than attacker 2, only sharing the trigger at different time steps. Thus, the model learns a more complex insight than a single device setup. Using two coordinated attackers results in better attacks than in a single attacker scenario.

\begin{table}[!htb]
\centering
\caption{ASR (\%) of the Time Bandits attack under different settings. $\epsilon = 0.1$.}
\label{tab:multiple_attk_2}
\resizebox{\linewidth}{!}{%
\begin{tabular}{cccccccccc}
\toprule
          & \multicolumn{3}{c}{10} & \multicolumn{3}{c}{25} & \multicolumn{3}{c}{50} \\
          & 0.1    & 0.5    & 1    & 0.1    & 0.5    & 1    & 0.1    & 0.5    & 1    \\ \midrule
N-MNIST   & 100 & 100 & 100 & 4.61 & 100 & 100 & 100 & 100 & 100   \\
N-CIFAR10 & 49.00 & 99.70 & 100 & 11.45 & 100 & 86.60 & 55.55 & 40.45 & 60.77 \\
Gesture   & 28.24 & 13.31 & 11.81 & 7.99 & 19.79 & 26.04 & 6.13 & 29.17 & 32.29  \\ \bottomrule
\end{tabular}%
}
\end{table}

The stealthiness of the backdoor samples in the multiple attackers scenario is also improved compared to the single attacker scenario since fewer time steps are modified. We evaluate the attacks' stealthiness in~\autoref{sec:stealthiness}. Overall, using Time Bandits compared to a single attacker setup (even when using a scaling factor) generally improves the attack performance and stealthiness. Similarly, the Time Bandits attack grants the attacker more flexibility and control over the trigger, which could be adapted to bypass known defenses or constraints.

\section{Ablation Study}

We first evaluate the attack performance by increasing the number of coordinated attackers. Then, we evaluate how well the clean network and the network under attack operate when the data distribution is non-IID.

\subsection{Effect of the Number of Attackers}
\label{subsec:multiple-attackers}

Now, we consider the effect of different amounts of malicious devices for our Time Bandits attack. Precisely, we split the trigger into 4 and 8 malicious devices. \autoref{tab:multiple_attk_5_10} shows the \ac{ASR} of different attack setups, with 4 and 8 malicious devices. 

\begin{table}[]
\centering
\caption{ASR (\%) of the Time Bandits attack under different settings. 4/8 attackers. $\epsilon = 0.1$.}
\label{tab:multiple_attk_5_10}
\resizebox{\linewidth}{!}{%
\begin{tabular}{ccccccc}
\toprule
          & \multicolumn{3}{c}{25} & \multicolumn{3}{c}{50} \\
          & 0.1    & 0.5    & 1    & 0.1    & 0.5    & 1    \\ \midrule
N-MNIST   & 100/100 & 100/100 & 100/100 & 7.86/7.49 & 100/50.16 & 100/100    \\
N-CIFAR10 & 56.60/54.65 & 99.95/99.90 & 100/99.40 & 65.65/99.95 & 40.70/55.85 & 100/100 \\ 
Gesture   & 25.17/42.19 & 52.78/69.27 & 59.72/76.04 & 67.53/59.03 & 77.08/73.26 & 55.38/84.38 \\ \bottomrule
\end{tabular}%
}
\end{table}

We observe interesting findings: first, the more attackers there are, the better the attack performance. This is mainly a cause of the probability of an attacker being selected. Note that the number of triggers in a sample is always the same, regardless of the number of attackers. In some cases (N-MNIST with 50 devices and half of them selected on each round), the attacker is not chosen in the latter rounds, making the attack less practical. However, overall, the attack is boosted by the number of attackers when the fraction of devices selected is small. In setups where all the devices are selected in each round, there is a slight difference in most cases, solely with the Gesture dataset, as the dataset is small. We conclude that more attackers are useful when a few devices are selected in each round or when the stealthiness must be improved. Further details are in~\autoref{sec:stealthiness}.

\subsection{Effect of the IID-ness}
\label{sec:non-iid}

In addition to the number of devices in the network, other properties, such as the data distribution, also affect the performance of the \ac{FL} network. This section evaluates how \acp{SNN} and the attack perform as the data distribution becomes non-IID. We systematically create the non-IID data using the Dirichlet distribution, a standard method in \ac{FL}~\cite{zhu2021federated}. We set the non-IID degree to 50\%. 
We first consider training solely on clean data to compare its performance with the IID version. Based on the results (see~\autoref{tab:baseline_fl_non_iid}), we observe a slight degradation in the accuracy. However, the degradation is not severe and also happens in regular non-spiking networks when non-IID data is used~\cite{venkatesha2021federated}. The results for different degrees of IID-ness, are in~\autoref{app:results}. 

\begin{table}[!htb]
\centering
\caption{Clean accuracy (\%) baseline for different FL setups under non-IID settings.}
\label{tab:baseline_fl_non_iid}
\resizebox{\columnwidth}{!}{%
\begin{tabular}{@{}cccccccccc@{}}
\toprule
       & \multicolumn{3}{c}{10} & \multicolumn{3}{c}{25} & \multicolumn{3}{c}{50} \\
 & 0.1    & 0.5    & 1    & 0.1    & 0.5    & 1    & 0.1    & 0.5    & 1    \\ \midrule
N-MNIST   &  89.18 & 98.31 & 98.17 & 95.31 & 97.31 & 97.14 & 92.90 & 95.94 & 96.11   \\
N-CIFAR10 &  38.05 & 45.60 & 48.30 & 21.75 & 23.65 & 26.70 & 17.30 & 20.35 & 18.30   \\
Gesture   &   47.05 & 70.31 & 72.92 & 44.62 & 56.25 & 56.08 & 42.71 & 49.13 & 49.83  \\ \bottomrule
\end{tabular}%
}
\end{table}

Regarding the attack with two attackers and non-IID settings, see~\autoref{tab:attack_fl_non_iid}, the attack performance drastically lowers on complex datasets. We achieve a maximum \ac{ASR} of 50.87\% for the Gesture dataset. This is primarily due to the dataset being complex and small. Thus, the attackers have few samples, which are non-IID and less diverse than in IID. This is not noticed for N-MNIST since the dataset is simpler and more extensive. However, we also notice a low \ac{ASR} on N-MNIST with 25 devices. This cause has also been observed previously in~\autoref{tab:single_attk}, and it is due to the selection of the attacker. Since only 10\% of the devices are selected in each epoch, the attackers might not be selected for some rounds (or none more), vanishing the backdoor effect. This effect is also linked to at what round an attacker is selected because the backdoor is better injected at the last rounds~\cite{Bagdasaryan2020}.

\begin{table}[htb]
\centering
\caption{ASR (\%) for different FL setups under non-IID settings, using Time Bandits with two attackers.}
\label{tab:attack_fl_non_iid}
\resizebox{\columnwidth}{!}{%
\begin{tabular}{@{}cccccccccc@{}}
\toprule
       & \multicolumn{3}{c}{10} & \multicolumn{3}{c}{25} & \multicolumn{3}{c}{50} \\
 & 0.1    & 0.5    & 1    & 0.1    & 0.5    & 1    & 0.1    & 0.5    & 1    \\ \midrule
N-MNIST & 86.32 & 100 & 100 & 55.13 & 100 & 100 & 100 & 100 & 100   \\
N-CIFAR10 & 8.45 & 100 & 100 & 6.30 & 58.30 & 81.65 & 12.20 & 24.20 & 73.95 \\
Gesture  & 4.51 & 21.18 & 14.41 & 30.73 & 19.79 & 50.87 & 29.34 & 25.00 & 22.92 \\ \bottomrule
\end{tabular}%
}
\end{table}

Interestingly, when using non-IID data over a small dataset (as Gesture), the attacker will not share samples from the same class. This causes an interesting effect that when injecting the trigger, the attacker uses both distinct time steps and distinct classes, which hardens even more to include the backdoor task in the model, see~\autoref{tab:multiple_attk_5_10_non_iid}. However, the backdoor is correctly injected in the model when the dataset is large and simpler, i.e., N-MNIST.

\begin{table}[htb]
\centering
\caption{ASR (\%) of Time Bandits attack under different settings with non-IID data. 4/8 attackers. $\epsilon = 0.1$.}
\label{tab:multiple_attk_5_10_non_iid}
\resizebox{\linewidth}{!}{%
\begin{tabular}{ccccccc}
\toprule
          & \multicolumn{3}{c}{25} & \multicolumn{3}{c}{50} \\
          & 0.1    & 0.5    & 1    & 0.1    & 0.5    & 1    \\ \midrule
N-MNIST   & 100 / 100 & 100 / 100 & 100 / 100 & 100 / 9.3 & 100 / 100 & 100 / 100   \\
N-CIFAR10 & 8.65 / 19.4 & 16.55 / 36.0 & 73.55 / 49.1 & 0.00 / 24.2 & 31.90 / 36.8 & 57.85 / 75.9  \\
Gesture   & 3.47 / 78.5 & 28.12 / 47.6 & 50.69 / 38.5 & 6.25 / 40.6 & 21.88 / 67.4 & 18.06 / 25.7 \\ \bottomrule
\end{tabular}%
}
\end{table}

\section{Defense Evaluation}

Numerous defenses have been designed against backdoor attacks. Specifically in \ac{FL}, the defenses can be applied by i) the local device and ii) by the central aggregator. None of the defenses in the literature are specific to \acp{SNN} or neuromorphic data. Thus, they must be adapted to handle neuromorphic data, \acp{SNN}, or \ac{FL}.

We consider \ac{STRIP}~\cite{gao2019strip} as a defense deployed locally by the devices, which measures the entropy of the clean and backdoor samples to detect outliers. This defense could be deployed by a device locally at inference time to reject inputs that have an abnormal entropy level, i.e., contain a trigger. Other methods, such as \ac{ABS}~\cite{liu2019abs}, are ineffective against \acp{SNN} with neuromorphic data~\cite{abad2023sneaky}.
\Ac{STRIP} considers access to a clean test set. First, by superimposing clean samples, \ac{STRIP} determines a detection boundary. These are then fed into the model, whose outputs are used to measure the Shannon entropy. Then, \ac{STRIP} assumes that the outputs should follow a normal distribution, from which it extracts the probability density function, which will serve as the detection boundary. Thus, a sample will be superimposed by a random \textbf{clean} sample. The perturbed sample's entropy below the boundary is considered clean or will be flagged as malicious otherwise.

\begin{figure*}[htb]
    \centering
     \begin{subfigure}[b]{0.16\linewidth}
        \includegraphics[width=\linewidth]{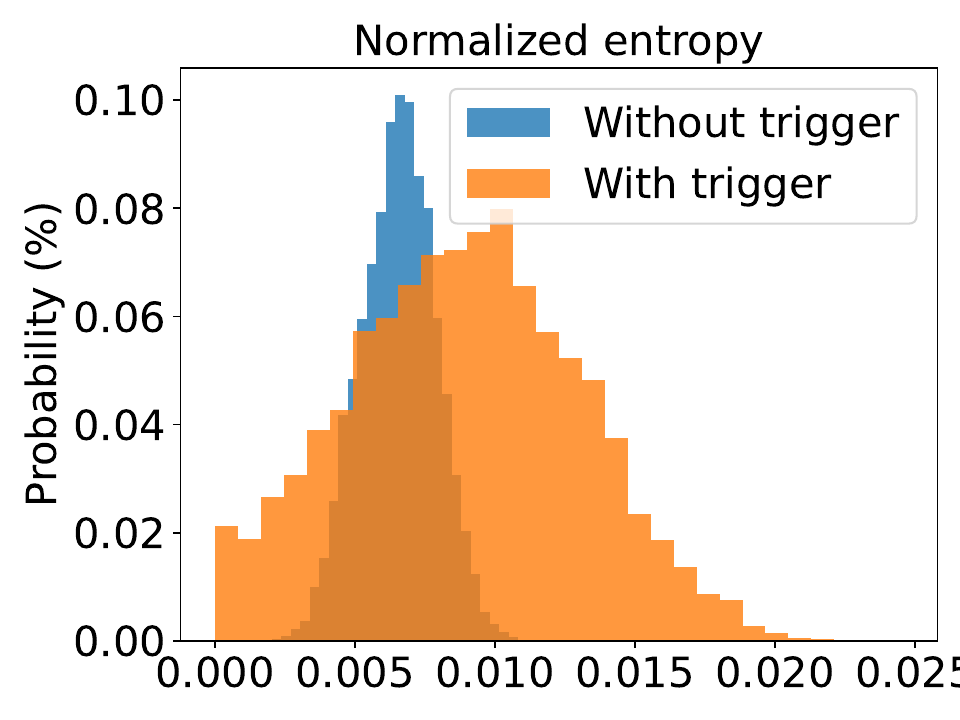}
    \end{subfigure}
     \begin{subfigure}[b]{0.16\linewidth}
        \includegraphics[width=\linewidth]{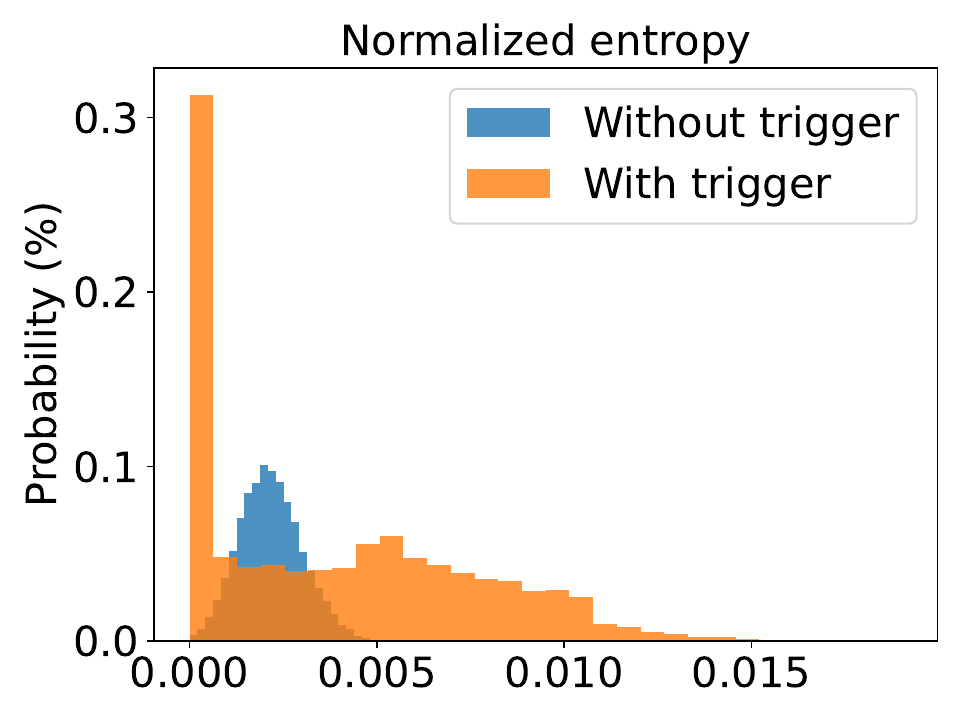}
    \end{subfigure}
     \begin{subfigure}[b]{0.16\linewidth}
        \includegraphics[width=\linewidth]{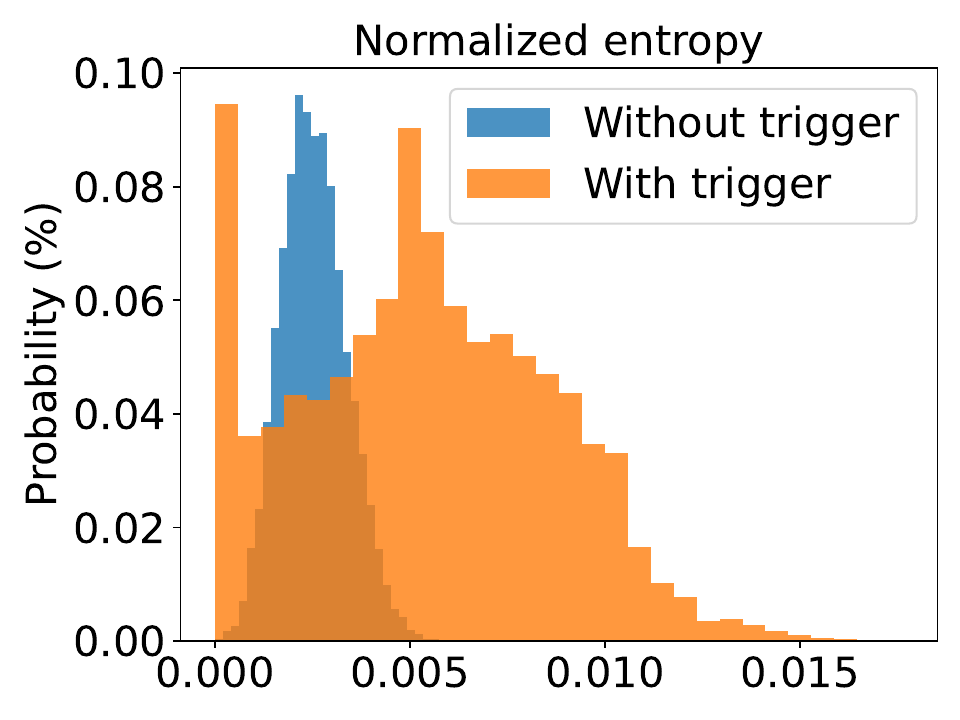}
    \end{subfigure}
    \begin{subfigure}[b]{0.16\linewidth}
        \includegraphics[width=\linewidth]{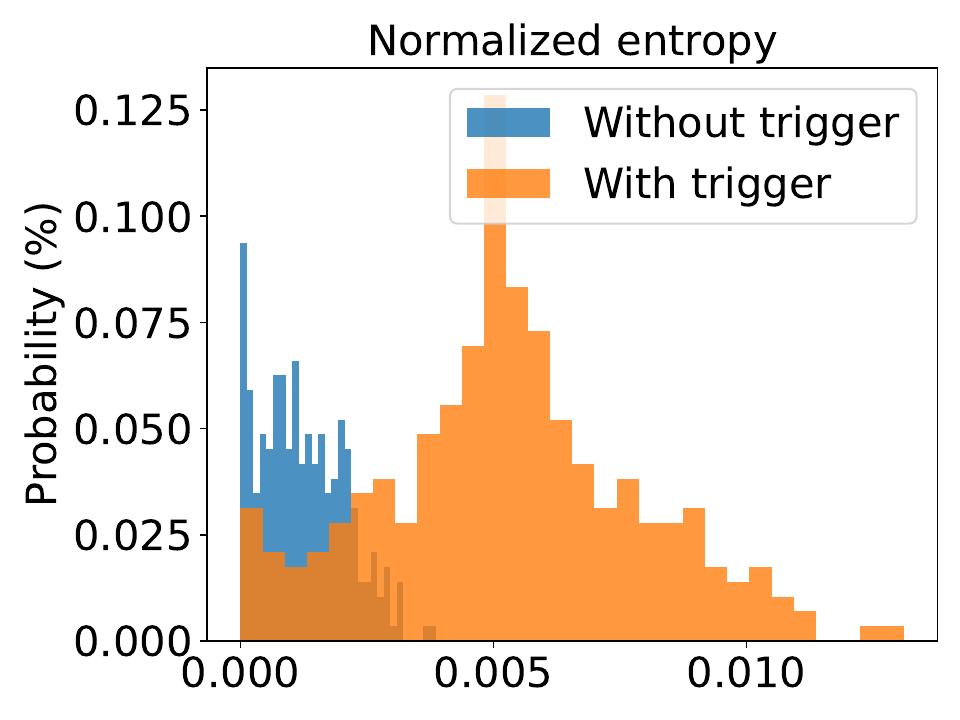}
    \end{subfigure}
    \begin{subfigure}[b]{0.16\linewidth}
        \includegraphics[width=\linewidth]{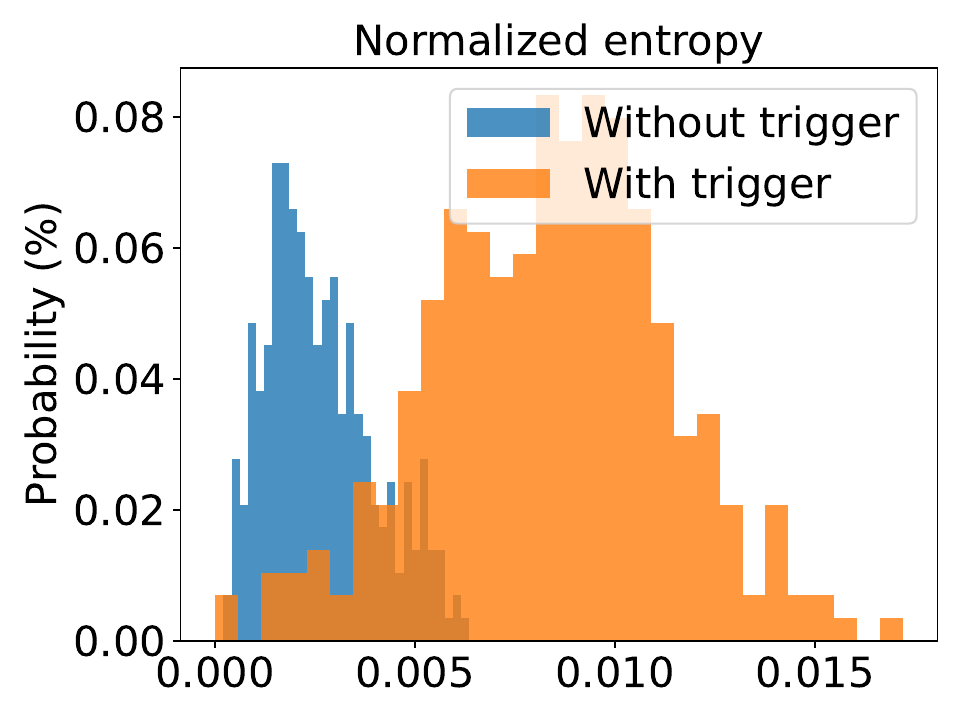}
    \end{subfigure}
    \begin{subfigure}[b]{0.16\linewidth}
        \includegraphics[width=\linewidth]{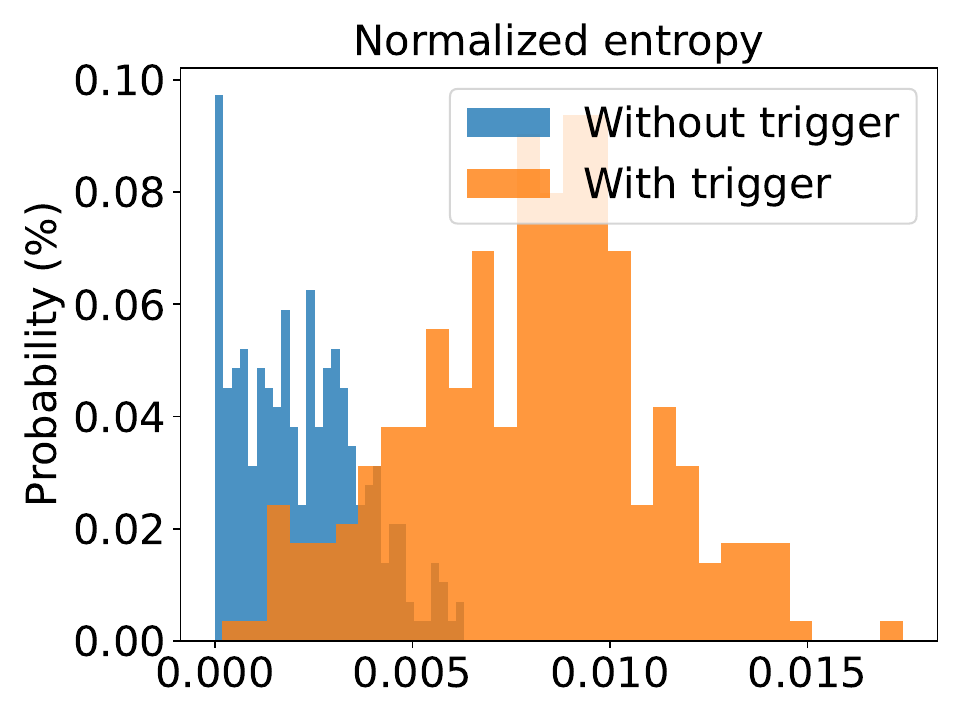}
    \end{subfigure}

     \begin{subfigure}[b]{0.16\linewidth}
        \includegraphics[width=\linewidth]{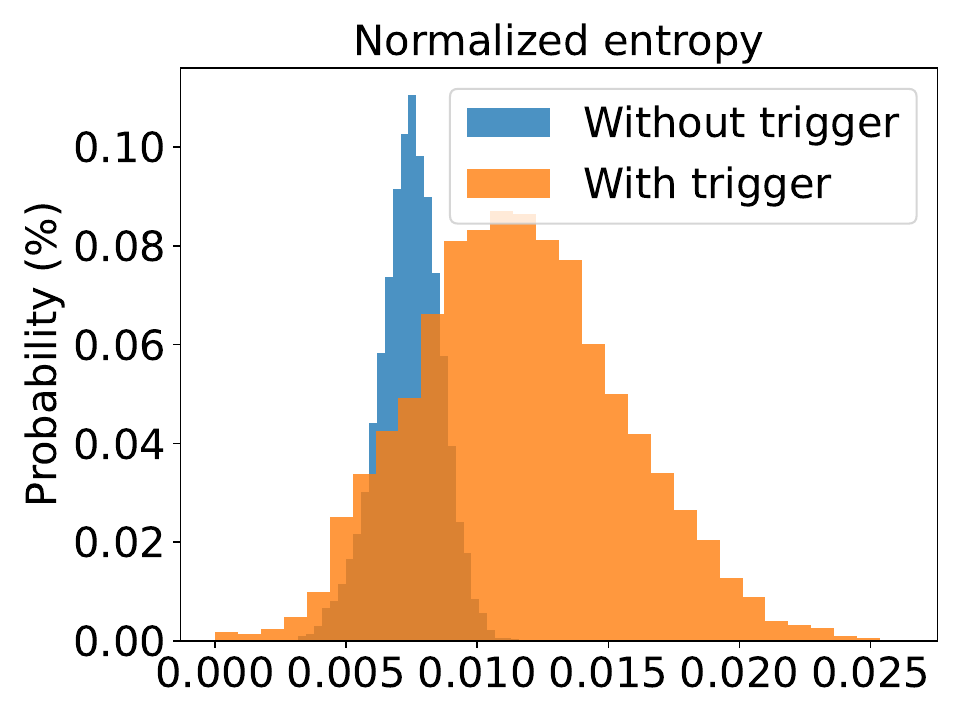}
        \subcaption{N-MNIST 10/0.1.}
    \end{subfigure}
     \begin{subfigure}[b]{0.16\linewidth}
        \includegraphics[width=\linewidth]{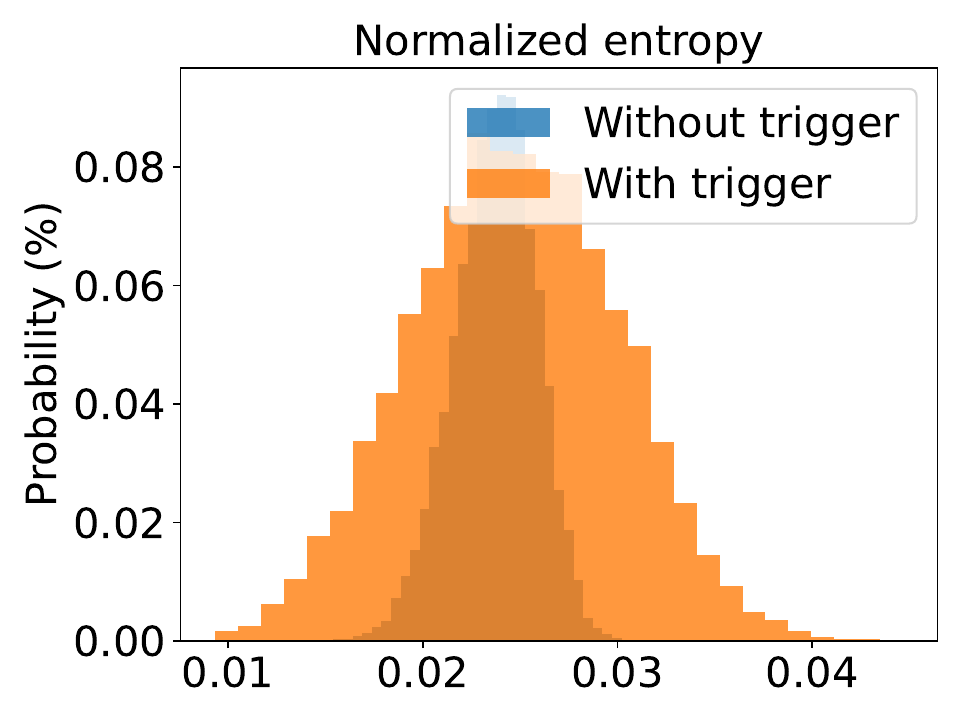}
        \subcaption{N-MNIST 25/0.1.}
    \end{subfigure}
     \begin{subfigure}[b]{0.16\linewidth}
        \includegraphics[width=\linewidth]{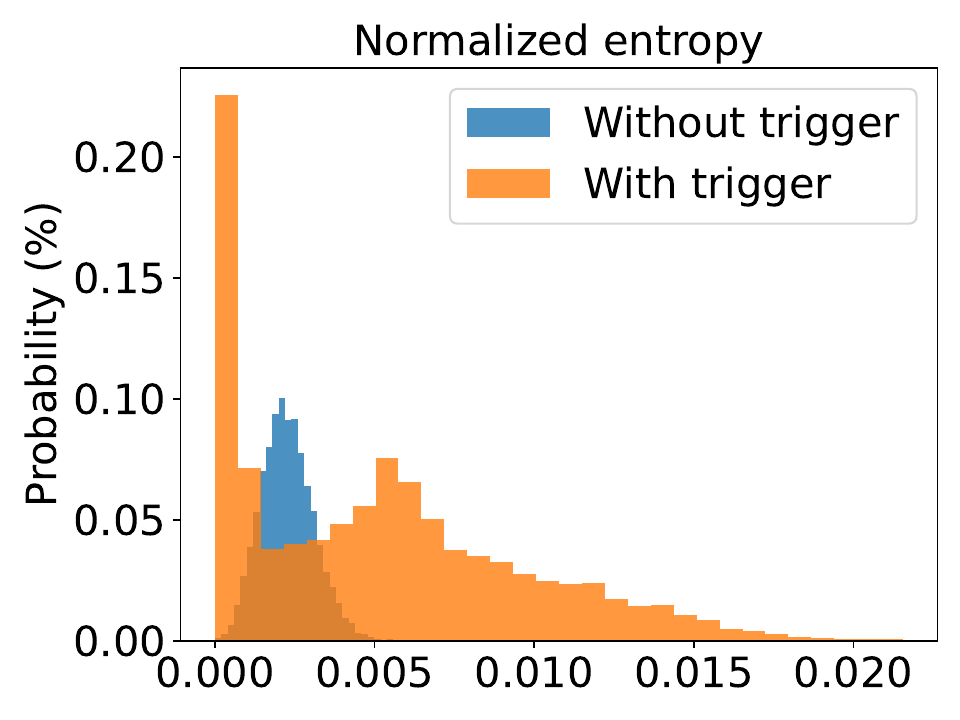}
        \subcaption{N-MNIST 50/0.1.}
    \end{subfigure}
    \begin{subfigure}[b]{0.16\linewidth}
        \includegraphics[width=\linewidth]{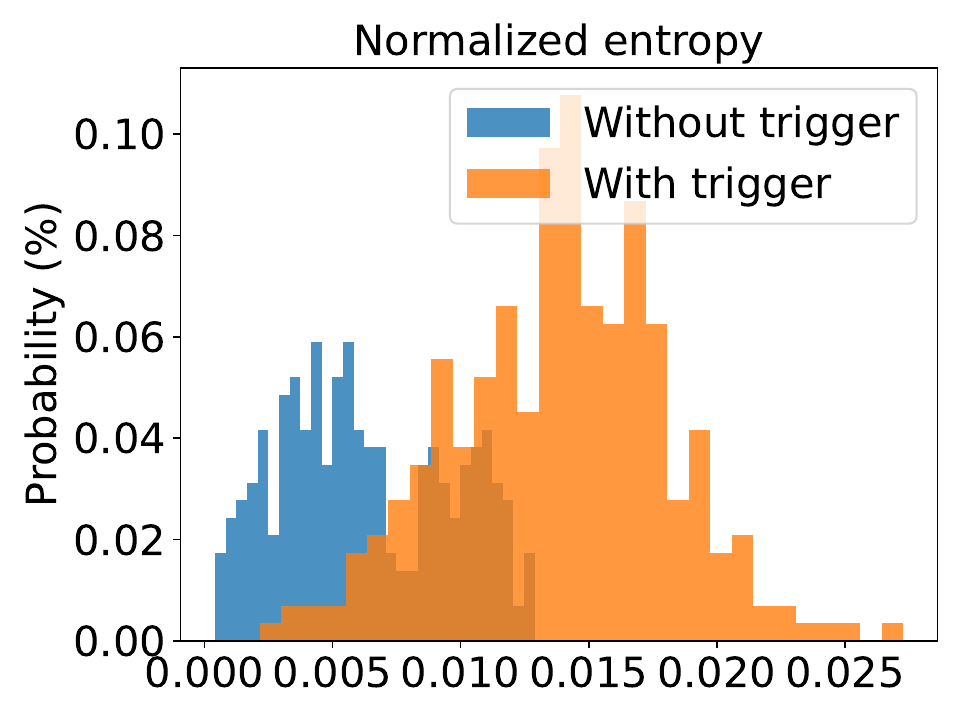}
        \subcaption{Gesture 25/0.1.}
    \end{subfigure}
    \begin{subfigure}[b]{0.16\linewidth}
        \includegraphics[width=\linewidth]{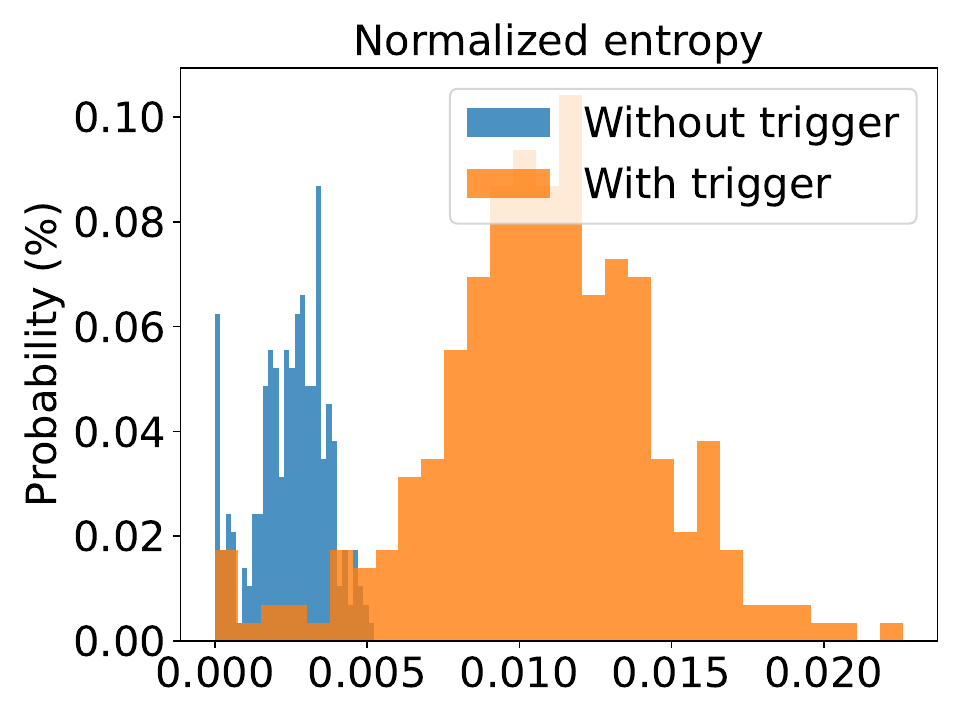}
        \subcaption{Gesture 25/0.5.}
    \end{subfigure}
    \begin{subfigure}[b]{0.16\linewidth}
        \includegraphics[width=\linewidth]{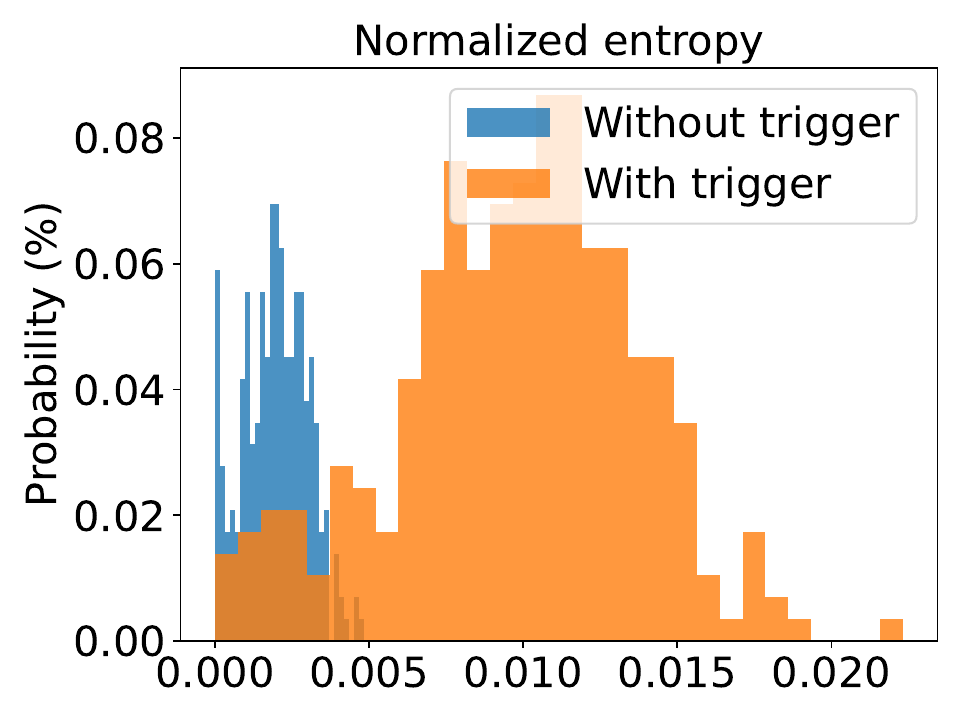}
        \subcaption{Gesture 25/1.0.}
    \end{subfigure}
    \caption{STRIP defense results for Time Bandits attack with different numbers/fractions of devices selected per epoch. The first row is non-IID, and the second row is IID.}
    \label{fig:strip}
\end{figure*}

To evaluate the effectiveness of \ac{STRIP}, we choose different settings where the backdoor is successful; we consider our Time Bandits attack under the N-MNIST and Gesture datasets (results for N-CIFAR10 are in Appendix B). To evaluate the effect of the number of devices and the fraction of devices selected in each round, we fixed the fraction of devices for the N-MNIST case. We fixed the number of total devices for the Gesture dataset to evaluate the effect of the selected number of devices. Based on the results (see~\autoref{fig:strip}), the entropy levels of clean and malicious samples are very similar. Then, it becomes challenging to find a reasonable boundary to separate clean and malicious samples. Moreover, when using IID data (bottom row), the malicious samples follow a similar entropy distribution to that in non-IID settings (top row).
Finally, as for Gesture, even the clean data does not follow a normal distribution of noisy and complex data. This makes it even more challenging to discard malicious samples. Likewise, in all the scenarios, the malicious samples overlap with the entropy levels of the clean data. 

\section{Stealthiness}
\label{sec:stealthiness}

Although there is no precise method for measuring the stealthiness of a trigger in backdoor attacks or the perturbation in adversarial examples, \ac{MSE} and \ac{SSIM} are commonly used. Following recent work~\cite{abad2023sneaky}, we also utilize these metrics for evaluating the stealthiness of the triggers.
\ac{MSE} compares the pixel-wise difference between the clean and the backdoored sample. \Ac{MSE} does not consider the context, e.g., the neighboring pixels, making its measurement local. \ac{SSIM} overcomes this limitation by considering the context of the image. To adapt these metrics to neuromorphic data, we compute them per frame and then average them. We also calculate the \ac{MSE} and \ac{SSIM} per batch, averaging the results. \autoref{tab:stealthiness} shows the average stealthiness of the proposed triggers.

\begin{table}[htb]
\centering
\caption{Average stealthiness of single attacker and Time Bandits attacks with a trigger size of 30\% of the source image.}
\label{tab:stealthiness}
\resizebox{\columnwidth}{!}{%
\begin{tabular}{ccccccc}
\hline
\multirow{2}{*}{\# of Attackers}   & \multicolumn{3}{c}{MSE $\downarrow$} & \multicolumn{3}{c}{SSIM $\uparrow$} \\
          & 1     &    2   &  3   &   1     &   2     &    3  \\ \hline
N-MNIST   &       8.57     &  2.14    & \textbf{0.84}     &   87.72     &    88.77    &    \textbf{90.01}   \\
Gesture   &      10.11      &   2.05   &  \textbf{0.98}    &    91.67    &   92.31     &    \textbf{92.97}   \\
N-CIFAR10 &       1.28     &   0.56   &  \textbf{0.13}    &   98.17     &   97.52     &    \textbf{98.88}   \\ \hline
\end{tabular}%
}
\end{table}

We observe that the triggers' stealthiness is high regardless of the number of attackers. Nevertheless, we notice that the more malicious devices in the network, the higher the stealthiness because fewer frames include the trigger. This empirical result is interesting because in the extreme case where the number of attackers is equal to the frame number, the poisoned image is left (almost) unchanged except for a single frame, boosting its stealthiness. 

\section{Related Work}

\Ac{FL} have been extensively used in many non-neuromorphic domains, in both \emph{general} \ac{DL} and \acp{SNN} setups. Specific to last,~\cite{venkatesha2021federated,wang2023efficient} evaluated the usability of \acp{SNN} in federated setups. Precisely, they experiment with a wide range of hyperparameters and non-neuromorphic datasets, e.g., N-MNIST, N-CIFAR10, or Gesture, showing that \ac{FL} is a suitable training scheme when in combination with \acp{SNN}. As stated, we are the first to test \ac{FL} with \acp{SNN} and neuromorphic data. 

Specific to the security of \acp{SNN},~\cite{xu2022securing,buchel2021adversarial,sharmin2019comprehensive} evaluated adversarial examples using both regular and neuromorphic data. On backdoor attacks, recent works have evaluated \acp{SNN} with neuromorphic data~\cite{abad2022poster,abad2023sneaky}, even with neuromorphic triggers that change with time, which are invisible to the human eye.

Regarding backdoor attacks in \ac{FL},~\cite{Bagdasaryan2020} explored distinct trigger injecting methods, showing that scaling the local model before sharing with the central server achieved better attack results. Consequent work~\cite{xie2020dba} considered a multiple attacker setup, where the attackers coordinate to split the trigger, where each malicious device uses its trigger part to train a malicious model locally. The central server aggregates the models, leading to launching the backdoor effect by using the entire trigger at test time.

On the defense side, \ac{FL} can launch defenses in different network parts, such as in the central server or by each device. On the device side, the devices may use defensive methods standard in \ac{DL}, such as \ac{ABS}~\cite{liu2019abs} or spectral signatures~\cite{tran2018spectral}. Still, these methods are ineffective with neuromorphic data~\cite{abad2023sneaky}. Some other defenses, which are \ac{FL} specific, can be deployed by the central server, e.g., Krum~\cite{blanchard2017machine} and FoolsGold~\cite{fung2020limitations}, which have not yet been explored in the realm of \acp{SNN}. Defenses deployed by the central server should be explored and adapted by considering neuromorphic data, which we will consider in future work.

\section{Conclusions \& Future Work}

Using neuromorphic data in IID or non-IID settings, we have demonstrated the feasibility of \ac{FL} with \acp{SNN}. We have also shown that increasing the number of devices in the network reduces its performance. Likewise, the fraction of devices selected in each round also affects the accuracy. These findings lead to the necessity of exploring the combination of \ac{FL}, \acp{SNN}, and neuromorphic data, improving its performance in realistic scenarios.

We first showed that the performance of existing (adapted) backdoor methods, i.e., BadNets, leads to a sub-optimal attack performance, decreasing performance when the number of devices increases. Compromising \acp{SNN} is more complicated than compromising standard \acp{DNN}. We then leverage scaling the attacker malicious model to increase its importance on the global averaged model, showing an improved attack performance. 
We developed a novel backdoor attack, i.e., Time Bandits, when multiple attackers collaborate in the \ac{FL} network. We showed that splitting the trigger through time and among the attackers makes the global model learn a complex link between the trigger and a backdoor task. This method improves the trigger stealthiness during training, which makes it more challenging to detect by human inspection.
We evaluated a state-of-the-art defense, \ac{STRIP}, which discards malicious samples at test time. After adopting the defense for its usage with neuromorphic data and \acp{SNN}, we showed that clean and malicious samples present similar entropy levels; thus, differentiating between them is difficult.
Having shown the vulnerability of \ac{FL} with \acp{SNN} and neuromorphic data and its growing interest, we emphasize the necessity for deeper exploration of neuromorphic-tailored defense mechanisms. 

\bibliographystyle{plain}
\bibliography{references}


\appendix
\section{Additional Experiments on IID-nes}
\label{app:results}

Apart from evaluation with IID and non-IID data (with a non-IID degree of 50\%), we additionally consider more degrees of non-IIIDness to explore the effect of it for the single attacker (using the scaling attack) and the Time Bandits attack. Using a non-IID degree of 0 equals a complete IID distribution. Similarly, a non-IID degree of 1 is a complete non-IID setup. We consider N-MNIST as a use case, where we vary the degree of non-IIDness from 0 (prior results in~\autoref{tab:multiple_attk_2}) to 75\%. According to the results (see~\autoref{tab:non-iid_degree}), we observe the potential of the Time Bandits attack. In most of the cases Time Bandits is necessary to achieve a successful attack.

\begin{table}[htb]
\centering
\caption{Scaled attacks with a single attacker and Time Bandits with two attackers for different degrees of non-kindness and datasets, using N-MNIST. 0 means completely IID.}
\label{tab:non-iid_degree}
\resizebox{\linewidth}{!}{%
\begin{tabular}{ccccccccccc}
\toprule
                           &      & \multicolumn{3}{c}{10} & \multicolumn{3}{c}{25} & \multicolumn{3}{c}{50} \\
                           &      & 0.1    & 0.5    & 1    & 0.1    & 0.5    & 1    & 0.1    & 0.5    & 1    \\ \midrule
                           & 0    &  39.98 / 100 & 100 / 100 & 100 / 100 & 39.41 / 4.61 & 100 / 100 & 100 / 100 & 39.36 / 100 & 100 / 100 & 100 / 100  \\
                           & 0.1  &  6.69 / 32.51 & 0.72 / 100 & 8.94 / 100 & 0.17 / 0.53 & 5.73                            / 100 & 9.08 / 100 & 9.86 / 64.23 & 21.92 / 57.83 & 11.97 / 99.33 \\
                           & 0.25 &  2.31 / 100 & 8.64 / 100 & 9.19 / 100 & 1.26 / 100 & 9.14 / 100 & 9.36 / 100 & 16.55 / 100 & 9.68 / 100 & 9.65 / 100   \\
                           & 0.75 &   10.29 / 99.51 & 9.56 / 100 & 9.61 / 100 & 7.95 / 100 & 8.53 / 100 & 9.39 / 100 & 9.78 / 100 & 9.18 / 100 & 9.14 / 100 \\ \bottomrule
\end{tabular}%
}
\end{table}

\section{Additional Defense Experiments}
\label{app:defense}

\begin{figure}[!h]
    \centering
     \begin{subfigure}[b]{0.32\linewidth}
        \includegraphics[width=\linewidth]{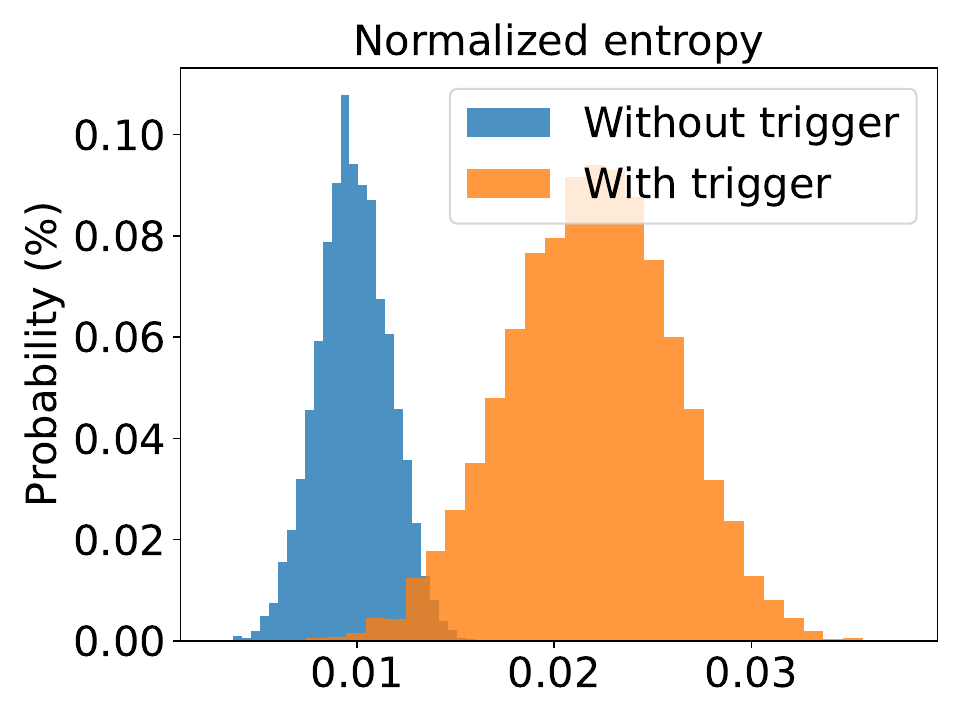}
    \end{subfigure}
     \begin{subfigure}[b]{0.32\linewidth}
        \includegraphics[width=\linewidth]{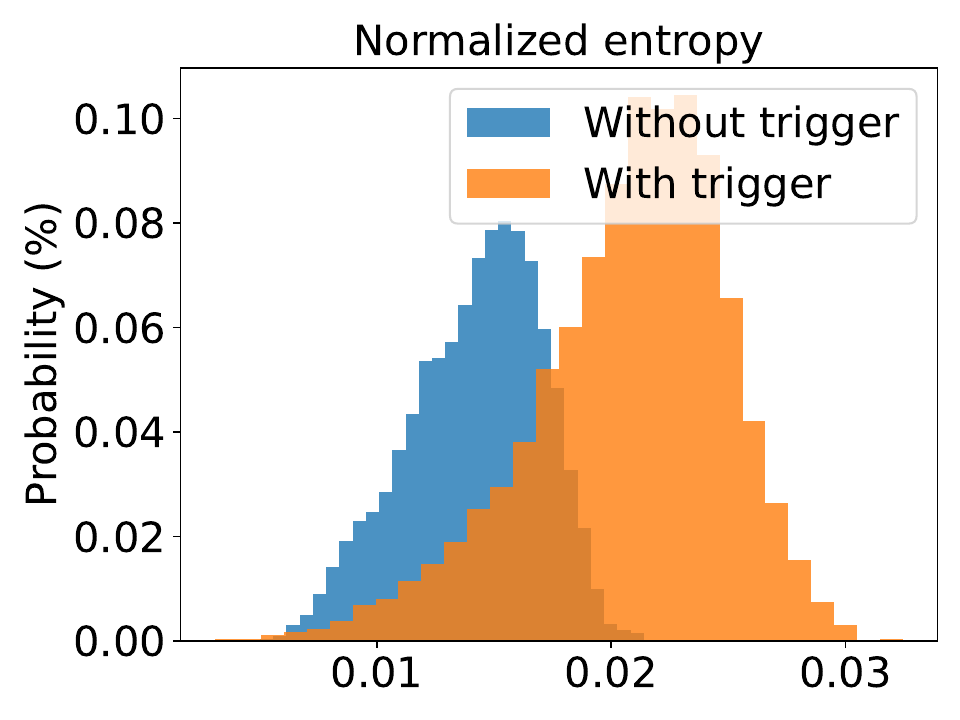}
    \end{subfigure}
     \begin{subfigure}[b]{0.32\linewidth}
        \includegraphics[width=\linewidth]{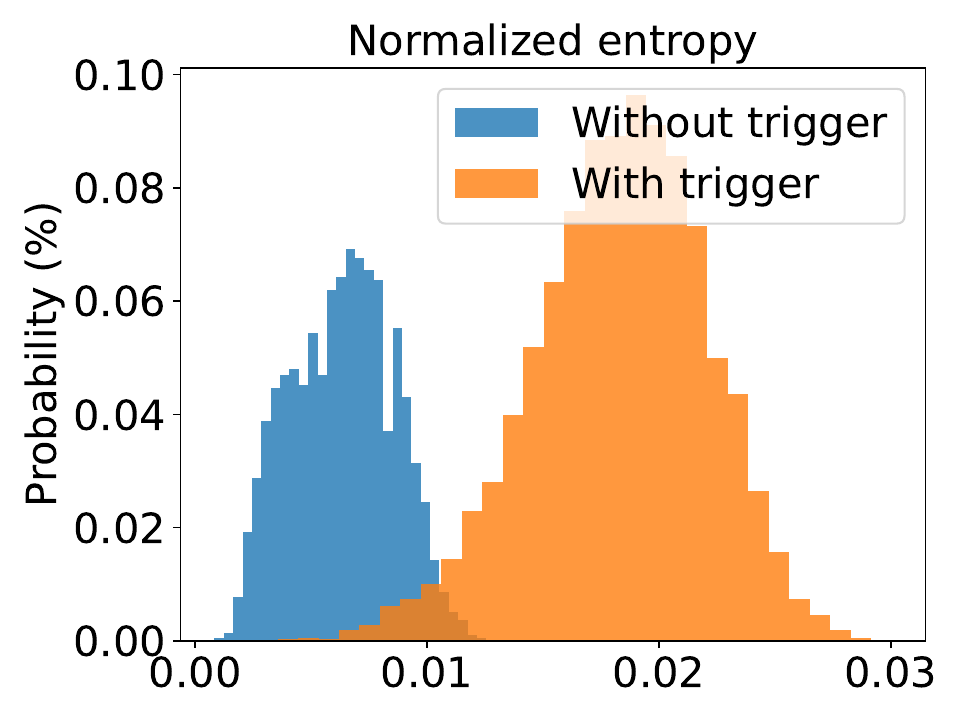}
    \end{subfigure}
    \begin{subfigure}[b]{0.32\linewidth}
        \includegraphics[width=\linewidth]{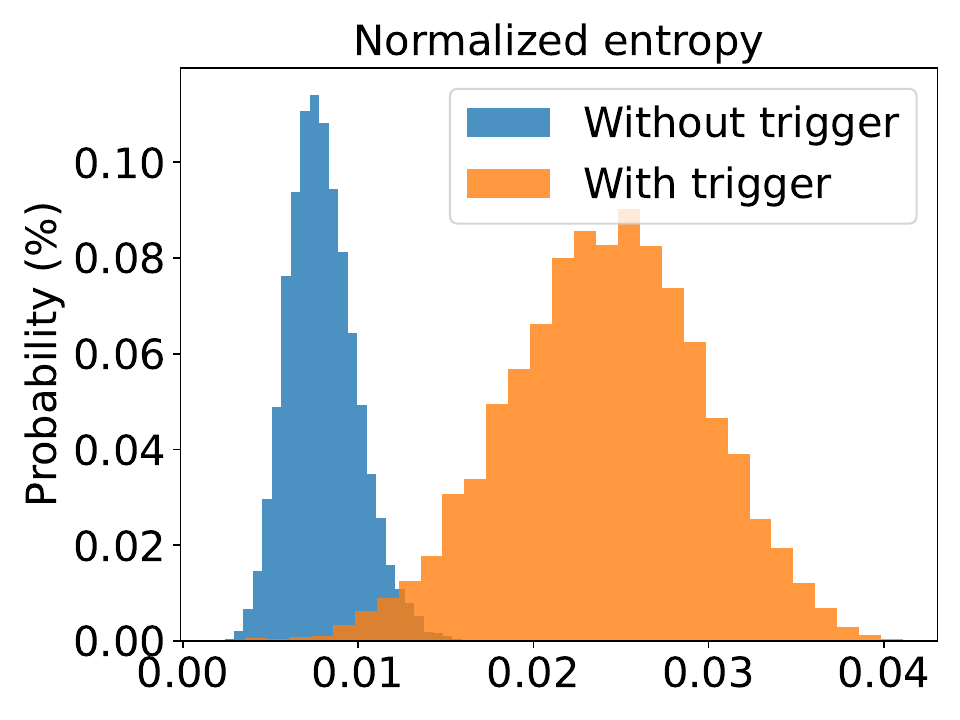}
        \subcaption{10/0.1.}
    \end{subfigure}
    \begin{subfigure}[b]{0.32\linewidth}
        \includegraphics[width=\linewidth]{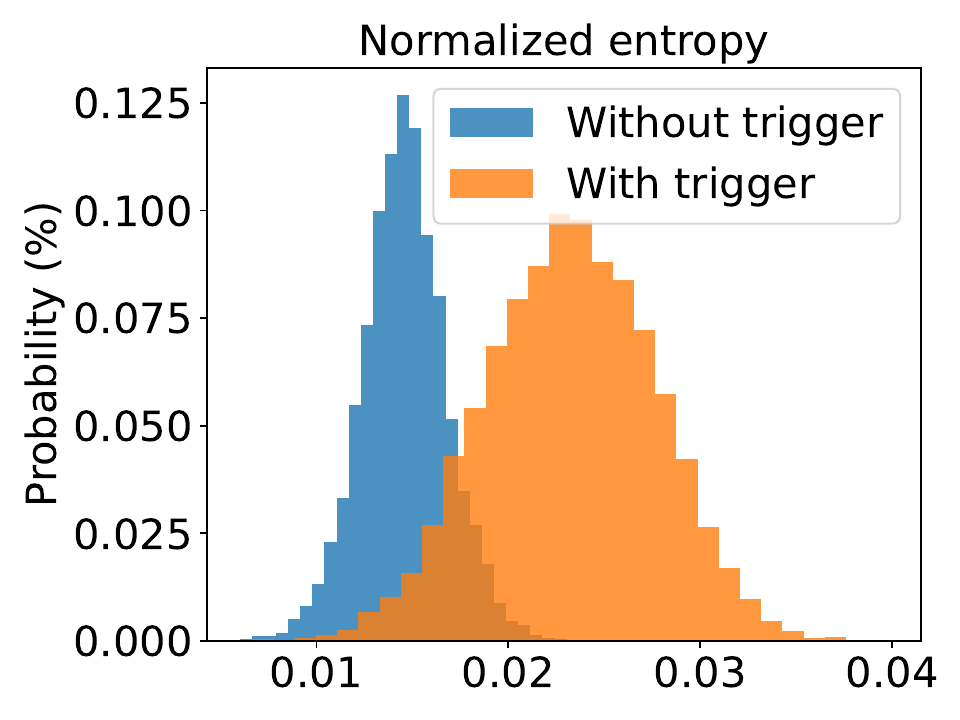}
        \subcaption{25/0.1.}
    \end{subfigure}
    \begin{subfigure}[b]{0.32\linewidth}
        \includegraphics[width=\linewidth]{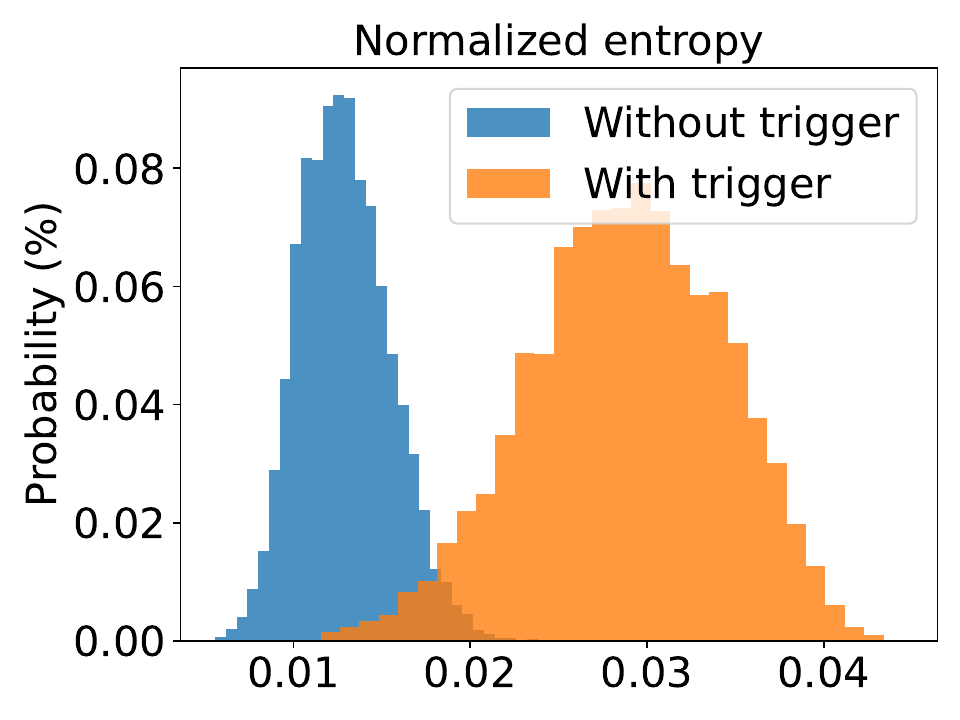}
        \subcaption{50/0.1.}
    \end{subfigure}
    \caption{STRIP defense results for Time Bandits attack with different numbers/fractions of devices selected per epoch, using N-CIFAR10. The first row is non-IID, and the second row is IID.}
    \label{fig:strip_cifar}
\end{figure}

In this section, we include the result for \ac{STRIP} on N-CIFAR10 for the Time Bandits attack with two malicious devices; see~\autoref{fig:strip_cifar}. The first row shows the normalized entropy for clean and poisoned non-IID data, and the bottom row considers IID data. Similar to another dataset, we observe that the clean and malicious data are similar. In all the cases, they overlap, separating between them. Nevertheless, N-CIFAR10 exhibits a slightly more apparent separation than in previously tested datasets. However, this separation is not enough and is not like the one observed by the authors in~\cite{gao2019strip} with non-neuromorphic data.

\end{document}